\documentclass[12pt]{article}
\usepackage{graphicx}

\newsavebox{\sboxpubnumber}
\newsavebox{\sboxpubdate}
\newcommand{\pubdate}[1]{\begin{lrbox}{\sboxpubdate}{#1}\end{lrbox}}

\newcommand{\Title}[1]{\begin{center} {\Large #1 } \end{center}}
\newcommand{\Author}[1]{\begin{center}{ \sc #1} \end{center}}
\newcommand{\Address}[1]{\begin{center}{ \it #1} \end{center}}

\newenvironment{Abstract}{\begin{quotation}  }{\end{quotation}}
\newenvironment{Presented}{\begin{quotation} \begin{center}
             PRESENTED AT\end{center}\bigskip
      \begin{center}\begin{large}}{\end{large}\end{center}
      \end{quotation}}

\begin{document}

\begin{titlepage}
\pubdate{\today}                    

\vfill
\Title{Warm Inflation Dissipative Dynamics}
\vfill
\Author{Arjun Berera\footnote{PPARC Advanced Fellow}}
\Address{Department of Physics and Astronomy, University of Edinburgh \\
         Edinburgh EH9 3JZ, United Kingdom}
\vfill
\vfill
\begin{Abstract}
Warm inflation dynamics is reviewed.  Results on dissipative scalar field
dynamics relevant to warm inflation are examined and estimates
of radiation production are given.
\end{Abstract}
\vfill
\begin{Presented}
    COSMO-01 \\
    Rovaniemi, Finland, \\
    August 29 -- September 4, 2001
\end{Presented}
\vfill
\end{titlepage}
\def\thefootnote{\fnsymbol{footnote}}
\setcounter{footnote}{0}


The inflation picture of accelerated scale factor expansion
has been long recognized as a potentially important
component of the early universe \cite{gs,si}, which
could solve the cosmological horizon and flatness problems.
However, consistent dynamical solutions of inflation founded
on quantum field theory have been allusive. Moreover, the thermodynamic
state during inflation is not unique.  In
the earliest conception of inflation \cite{gs,si,ni,ci}, it was
pictured that inflation would result in a isentropic
expansion that would rapidly put the universe in
a supercooled thermodynamic phase.  
Subsequently it was observed that supercooled
inflation is not mandatory and that nonisentropic 
inflationary expansion, warm inflation, also is
possible \cite{wi} (see also \cite{moss1}).
Representing the early universe by a two fluid mixture of
radiation energy density $\rho_r$ and vacuum energy density
$\rho_v$, the inflationary regime, when the scale factor 
accelerates ${\ddot R} > 0$, generally requires
$\rho_v > \rho_r$.  Thus from the point of view of
two fluid Friedmann cosmology, isentropic
inflationary expansion appears as a limiting case within the general regime
of nonisentropic inflation.

In particle physics the vacuum equation of state $\rho_v=-p_v$ is realized by a
scalar field with energy density 
$\rho(\phi) = {\dot \phi}^2/2 + (\nabla \phi)^2 /2 +V(\phi)$, in which the
potential energy density dominates 
\begin{equation}
V(\phi) \gg \frac{1}{2} {\dot \phi}^2, \frac{1}{2} (\nabla \phi)^2.
\label{cond}
\end{equation}
Most field theory descriptions of inflation represent the vacuum energy through
a scalar field satisfying Eq. (\ref{cond}), with $\phi$ referred to as the
inflaton.  The goal of inflationary scalar field dynamics is to sustain the
vacuum energy sufficiently long for expansion of the scale factor to exceed
observational lower bounds and then end the inflationary epoch by entering
the radiation dominated epoch.

The most nontrivial aspect of the inflaton models is understanding the
energy transfer dynamics from potential energy to radiation.
A commonly followed picture is that dissipative effects
of the inflaton field can be ignored throughout
the inflation period, thus leading to a supercooled inflationary
regime.  However, from a thermodynamic perspective, this
picture appears very restrictive. The point being, even if
the inflaton were to allow a minuscule fraction
of the energy to be released, say one part in $10^{20}$,
it still would constitute a significant radiation
energy density component in the universe.
{}For example, for inflation with vacuum (i.e. potential) energy
at the GUT scale $\sim 10^{15-16} {\rm GeV}$,
leaking one part in $10^{20}$ of this energy into radiation 
corresponds to a temperature of $10^{11} {\rm GeV}$,
which is nonnegligible.  In fact, the most relevant
lower bound that cosmology places on the temperature after inflation
comes from the success of hot Big-Bang nucleosynthesis,
which thus requires the universe to be within the
radiation dominated regime by $T \stackrel{>}{\sim} 1 {\rm GeV}$.
This limit can be met in the above example by dissipating 
as little as one part in $10^{60}$ of the vacuum energy
into radiation.
Thus, from the perspective of both interacting field theory and
basic notions of equipartition,  it appears to be a highly
tuned requirement of supercooled inflation to prohibit
the inflaton from even such tiny amounts of dissipation.

These considerations have led to extensive examination
of warm inflation.
Several types of
phenomenological warm inflation models exist
in the literature \cite{wi,wimodels}.  The presence
of radiation during inflation generally will affect the
seeds of primordial density fluctuations induced in
the inflaton field, which is a complicated two fluid problem.
The basic model assumes perfect thermalization \cite{bf2,abden}
and for such models observational tests have 
been developed \cite{tb,bh}.
These test show possibilities that
could differentiate warm inflation from supercooled
inflation.  
Warm inflation dynamics also has been studied in
quantum field theory \cite{bgr,yl,br,imoss} and some of the
phenomenological models can be derived from first 
principles \cite{abden,bgr2,bk}. 
Furthermore, conceptually there is a simplification
in the warm inflation picture 
in that the dynamics is
completely free of questions about quantum-to-classical
transition. The scalar inflaton field, both background
and fluctuation components \cite{bf2,abden}, is in a
classical state, thus immediately justifying the application
of  a classical evolution equation and implying that the 
induced metric perturbations are
classical.  

However despite the conceptual clarity and despite the suggestive
thermodynamic considerations, deriving this dynamics from first
principles quantum field theory is nontrivial.  The key reasons
primarily are technical.  To clarify this
point,  a comparison with supercooled inflationary
dynamics is useful.  In supercooled inflation,
the process of inflation and radiation production are
neatly divided into two different epochs, whereas in
warm inflation dynamics, both processes occur concurrently.
As such, for warm inflation dynamics there is considerable
and nontrivial interplay between the equations of background 
inflationary expansion and quantum field theory dynamics and thus it
become technically more difficult to solve than supercooled
inflation.  In effect, warm inflation solutions are of an
``all-or-nothing'' type in that if a solution works, it solves
everything and if something fails, the whole solution
becomes useless.  On the other hand,
supercooled inflation solutions are of a ``pick-and-choose''
type, in that every aspect of the problem is compartmentalized,
i.e. inflation, reheating, quantum-to-classical transition,
and there is little continuity amongst the different problems.

Statements have been made about the impossibility of warm inflation
dynamics \cite{yl}. However the dynamical considerations
leading up to these conclusions were limited in their scope,
as had been noted previous to this work \cite{bgr}.
In particular, these works looked for high temperature
warm inflation solutions, under rigid adiabatic, equilibrium
conditions.  Nevertheless, within this limited framework, 
one type of warm inflation solution
was obtained \cite{abden,bgr2,bk}, 
and due to the "all-or-nothing" nature
mentioned above, this solutions
can not be discarded as a serious contender
in any more complete theory of inflation \cite{bk}.
Moreover, these early works \cite{bgr,yl} have explicated
one very important point, that warm inflation dynamics is
not trivial and before it can be directly solved, several
missing gaps in the knowledge of dissipative dynamics
must be clarified.

As one step in this direction to fill the missing gaps, recently we
studied the zero temperature dissipative
dynamics of interacting scalar field 
systems in Minkowski spacetime \cite{br}.
This is useful to understand, since
the zero temperature limit constitutes a baseline effect that
will be prevalent in any general statistical state.
What our results show is that
for a broad range of cases, involving interaction
with as few as one or two fields, dissipative regimes are found
for the scalar field system.  This is important for inflationary
cosmology, since it suggests that dissipation
may be the norm not exception for an interacting scalar field system,
thus suggesting that warm
inflation could be a natural dynamics once proper treatment of
interactions is done.

Our analysis of dissipative dynamics starts with the general
Lagrangian,
\begin{eqnarray} 
{\cal L} [ \Phi, \chi_j, \bar{\psi}_k, \psi_k] &=&  
\frac{1}{2} 
(\partial_\mu \Phi)^2 - \frac{m_\phi^2}{2}\Phi^2 - 
\frac{\lambda}{4 !} \Phi^4  
+ \sum_{j=1}^{N_{\chi}} \left\{ 
\frac{1}{2} (\partial_\mu \chi_{j})^2 - \frac{m_{\chi_j}^2}{2}\chi_j^2 
- \frac{f_{j}}{4!} \chi_{j}^4 - \frac{g_{j}^2}{2} 
\Phi^2 \chi_{j}^2  
\right\} 
\nonumber \\ 
&+& \sum_{k=1}^{N_{\psi}}   
\bar{\psi}_{k} \left[i \not\!\partial - m_{\psi_k} -h_{k,\phi} \Phi 
- \sum_{j=1}^{N_\chi} h_{kj,\chi} \chi_j \right] \psi_{k} 
\: , 
\label{Nfields} 
\end{eqnarray} 
with $\Phi \equiv \varphi + \phi$ such that
$\langle \Phi \rangle = \varphi$.  Our aim is to
obtain the effective equation of motion for $\varphi(t)$ and
from that determine the energy dissipated from the $\varphi(t)$
system into radiation.

Using the tadpole method \cite{tadpole}, which requires 
$\langle \phi \rangle =0$,
the effective equation of motion for $\varphi(t)$ emerges
\begin{eqnarray} 
&&\ddot{\varphi}(t) + m_\phi^2 \varphi(t) + \frac{\lambda}{6} \varphi^3(t)  
+\frac{\lambda}{2} \varphi(t) \langle \phi^2 \rangle 
+\frac{\lambda}{6} \langle \phi^3 \rangle 
+\sum_{j=1}^{N_{\chi}} g_j^2 \left[\varphi (t) \langle \chi_j^2 \rangle + 
\langle \phi \chi_j^2 \rangle \right] \nonumber \\ 
&& + 
\sum_{k=1}^{N_{\psi}} h_{k,\phi} \langle \bar{\psi_k} \psi_k \rangle= 0 \;. 
\label{eq2phi1} 
\end{eqnarray} 
The field expectation values in this equation are
obtained by solving the coupled set of field equations.
In our calculation, we have evaluated them in a perturbative
expansion using dressed Green's functions \cite{br,lawrie1,bgr}. 
One general feature of these expectation values
is they will depend of the causal history of $\varphi(t)$,
so that Eq. (\ref{eq2phi1}) is a temporally nonlocal equation
of motion for $\varphi(t)$.  

The general expression for the effective equation
of motion is given in \cite{br} and is very complicated.
{}Formally, we can examine Eq. (\ref{eq2phi1})
within a Markovian-adiabatic approximation, in which
the equation of motion is local in time and the motion
of $\varphi(t)$ is slow.  At $T=0$, such an approximation
is not rigorously valid.  Nevertheless, this approximation
allows understanding the magnitude of dissipative effects.
{}Furthermore, we have shown in \cite{br} that the nonlocal
effects tend to filter only increasingly higher
frequency components of $\varphi(t)$
from nonlocal effects increasingly further back in time.  
Thus for low
frequency components of $\varphi(t)$, memory only is
retained to some short interval in the past.   Since within
the adiabatic approximation, $\varphi(t)$ only has low frequency 
components, we believe the Markovian-adiabatic
approximation is legitimate at least for order of
magnitude estimates. Within this approximation,
the effective equation of motion for $\varphi(t)$ has the general
form
\begin{eqnarray} 
&& \ddot{\varphi}(t) + m_{\phi}^2 \; \varphi (t) +  
\frac{\lambda}{6} \varphi^3 (t) + \eta (\varphi)  
\dot{\varphi} (t) =0\;, 
\label{final1} 
\end{eqnarray} 
where
\begin{eqnarray} 
&&\eta(\varphi)  =  
\varphi^2 (t) \frac{\lambda^2 \alpha_{\phi,\psi}^2 }{128 \pi \;  
\sqrt{m_{\phi}^4 +  
\alpha_{\phi,\psi}^4}\; \sqrt{2 \sqrt{m_{\phi}^4 + \alpha_{\phi,\psi}^4} + 
2m_{\phi}^2}}\nonumber \\ 
&& +  \frac{\lambda^2 \alpha_{\phi,\psi}^2}{4}  
\int \frac{d^3 {\bf q}_1}{(2 \pi)^3}  
\frac{d^3 {\bf q}_2}{(2 \pi)^3}  
\left\{ 
\frac{1}{\omega_\phi ({\bf q}_1)^2  
\omega_\phi ({\bf q}_2) \omega_\phi ({\bf q}_1+{\bf q}_2) 
\left[\omega_\phi ({\bf q}_1) + 
\omega_\phi ({\bf q}_2) +\omega_\phi ({\bf q}_1+{\bf q}_2)\right]^3}  
+ {\cal O}\left(\frac{\Gamma_{\phi}^2}{\omega_{\phi}^2} \right) \right\} 
\nonumber \\ 
&& +  \varphi^2(t) \sum_{j=1}^{N_\chi} g_j^4  
\frac{\alpha_{\chi,\psi}^2}{32 \pi} 
\frac{1}{\sqrt{m_{\chi_j}^4 + \alpha_{\chi,\psi}^4}\;  
\sqrt{2 \sqrt{m_{\chi_j}^4 + \alpha_{\chi,\psi}^4} + 
2m_{\chi_j}^2}}  \nonumber \\ 
&& + \sum_{j=1}^{N_\chi} \frac{g_j^4 \alpha_{\chi,\psi}^2}{4} \!\! 
\int \frac{d^3 {\bf q}_1}{(2 \pi)^3}  
\frac{d^3 {\bf q}_2}{(2 \pi)^3}  
\left\{ 
\frac{1}{\omega_{\chi_j} ({\bf q}_1)^2  
\omega_\phi ({\bf q}_2) \omega_\phi ({\bf q}_1+{\bf q}_2) 
\left[\omega_{\chi_j} ({\bf q}_1) + 
\omega_\phi ({\bf q}_2) +\omega_\phi ({\bf q}_1+{\bf q}_2)\right]^3}  
\right. \nonumber \\ 
&& \left. + {\cal O}\left(\frac{\Gamma_{\chi_j}^2,\Gamma_{\phi}^2}{\omega^2} 
\right) \right\}, 
\label{eta2} 
\end{eqnarray} 
 
\noindent 
with
\begin{equation} 
\alpha_{\phi,\psi}^2 = \sum_{k=1}^{N_\psi} \frac{h_{k}^2}{8 \pi} m_{\phi}^2 
\left(1-\frac{4 m_{\psi_k}^2}{m_{\phi}^2} \right)^{\frac{3}{2}} \;,
\label{alpha} 
\end{equation} 
and 
\begin{equation} 
\alpha_{\chi,\psi}^2 = \sum_{k=1}^{N_\psi} \frac{h_{kj,\chi}^2}{8 \pi}  
m_{\chi_j}^2 
\left(1-\frac{4 m_{\psi_k}^2}{m_{\chi_j}^2} \right)^{\frac{3}{2}} \;. 
\label{alpha2} 
\end{equation} 

As an alternative to the above Lagrangian based derivation,
a canonical derivation also can be attempted based on the formalism
developed by Morikawa and Sasaki in the mid 80's \cite{ms1}.  Although
the canonical and Lagrangian approaches should yield the same
final answer, the former is far less developed in dissipative
quantum field theory, in particular
for treating interactions.  Nevertheless, the canonical
approach provides useful insight, especially for understanding
the origin of particle creation.  For example, consider
the $\langle \chi_j^2 \rangle$ terms in Eq. (\ref{eq2phi1}).
In the canonical approach this expectation value is
expressed as
\begin{eqnarray}
\langle \chi_j^2 \rangle =
\int \frac{d^3 q}{(2 \pi)^3 2 {\rm Re} \omega_{\bf q,\chi_j}}
\left[2 x_{\bf q,\chi_j}(t) + 2 {\rm Re} y_{\bf q,\chi_j}(t)+1\right] ,
\label{averchi}
\end{eqnarray}
where
$x_{{\bf q},\chi_j}(t)= \langle a_{{\bf q},\chi_j}^{\dagger}(t) 
a_{{\bf q},\chi_j}(t) \rangle$
is the particle
number density and
$y_{\bf q,\chi_j}(t)= \langle a_{{\bf q},\chi_j}(t) a_{-{\bf q},\chi_j}(t) 
\rangle$ is the
off-diagonal correlation.  The evolution equations for
$x_{\bf q,\chi_j}(t)$ and $y_{\bf q,\chi_j}(t)$ can be obtained from the
field equation for $\chi_j$ to give
\begin{eqnarray}
&&{\dot x}_{{\bf q},\chi_j} = \frac{\dot{\omega}_{\bf q,\chi_j}}
{\omega_{\bf q,\chi_j}}
{\rm Re} \, y_{\bf q,\chi_j} \;,
\nonumber \\
&& \dot{y}_{\bf q,\chi_j}= \frac{\dot{\omega}_{\bf q,\chi_j}}
{\omega_{\bf q,\chi_j}}
\left[ x_{\bf q,\chi_j} +\frac{1}{2} \right] - 2 i 
\omega_{\bf q,\chi_j}
y_{\bf q,\chi_j} \;.
\label{diff}
\end{eqnarray}
To yield dissipation, it is noted in \cite{ms1} that 
the correlation
amongst produced particles needs to be destroyed
sufficiently rapidly and ${\dot x}_{\bf q,\chi_j}(t)$ and
$\langle \chi_j(t) \rangle$ should become local functions of time
involving $\varphi(t)$ and ${\dot \varphi}(t)$.
Based on this requirement,
\cite{ms1} assert that $\omega_{\bf q,\chi_j}$ in
the above equation for ${\dot y}_{\bf q,\chi_j}(t)$
must have an imaginary part, which in fact should be
the $\chi_j$-particle decay width,
${\rm Im} \omega_{{\bf q},\chi_j} \propto \Gamma_{\bf q,\chi_j}$.
Applying these assumptions, the contribution to
$\eta(\varphi)$ in Eq. (\ref{final1}) from the $\chi_j$ field, say
$\eta_{\chi_j}(\varphi)$, once again is obtained
(up to O(1) factors). 
Since there are ad-hoc assumptions necessary, this approach still
is incomplete and requires further development. 
Nevertheless, the approach is interesting and
for the time being accepting the assumptions,
the origin of particle creation and energy conservation
can be seen very clearly. In particular, the particle
production rate is given by
$\int (d^3k/(2\pi)^3){\dot x}_{\bf q,\chi_j}(t)\omega_{\bf q,\chi_j}$
and analogous to the example in \cite{ms1}, it can be shown
this is exactly equal to the vacuum energy loss rate
from the $\chi_j$ field contribution,
$\eta_{\chi_j}(\varphi) {\dot \varphi}^2$.

Returning to equation (\ref{final1}), some estimated magnitudes
of energy production will be obtained 
here with full details given in \cite{br}.  Our primary
interest is in the overdamped regime 
\begin{equation} 
m^2(\phi) = m_{\phi}^2 + \lambda \varphi^2/2 < \eta^2,
\end{equation} 
since this is the regime ultimately of interest to warm inflation.
In this regime,
the energy dissipated by the scalar field goes into  
radiation energy density $\rho_r$ at the rate
\begin{equation} 
{\dot \rho}_r = -\frac{dE_{\phi}}{dt} = 
\eta(\varphi) {\dot \varphi}^2 .
\end{equation} 

In \cite{br} we have determined radiation production for
two cases 
\begin{eqnarray}
{\rm (a). } \hspace{0.1cm} & & m(\varphi) > m_{\chi} > 2m_{\psi} \nonumber\\
{\rm (b). } \hspace{0.1cm} & & m_{\chi} > 2 m_{\psi} > m(\varphi).  
\end{eqnarray}
To focus on
a case typical for inflation, suppose the potential
energy is at the GUT scale $V(\varphi)^{1/4} \sim 10^{15} {\rm GeV}$
and we consider the other parameters in a regime consistent
with the e-fold and density fluctuation
requirements of inflation.  Note, although this is a flat nonexpanding
spacetime analysis, since the dissipative effects will
be at subhorizon scale, one expects these estimates to give a reasonable
idea of what to expect from a 
similar calculation done in expanding spacetime.
Expressing the radiation in terms of a temperature scale as
$T \sim \rho_r^{1/4}$, we find for case (a)
$1 {\rm GeV} < T < 10^7 {\rm GeV} < H$ and for case (b)
$T \stackrel{>}{\sim} 10^{14} {\rm GeV} > H$,
where $H = \sqrt{8 \pi V/(3m_p^2)}$.

It should be clarified that the results discussed here do not
require supersymmetry, although they
easily could be applied in SUSY models.  
{}For such models, the low-$T$ warm inflation 
solutions suggested by case (a)
could be useful in avoiding gravitino overproduction \cite{tl}.
Although, as an aside, even high temperature warm inflation
solutions could avoid gravitino overproduction
by thermal inflation type mechanisms \cite{ti}, in
which the inflaton is placed in a metastable phase
after warm inflation, and the universe inflates
isentropically, thus allowing the gravitinos to dilute.
Turning to case (b), in general this case seems more
interesting, since it offers a very robust
possibility for radiation production. 
This is what the formal calculation indicates.
Moreover, the canonical approach suggests a clear physical picture,
that as the background field $\varphi(t)$ evolves, it changes
the mass of the $\chi_j$-bosons, so that their positive and negative
frequency components mix.  This in turn results in the
coherent production of $\chi_j$ particles, which then rapidly decohere
through decay into lighter $\psi_k$-fermions.
This process appears to yield
robust warm inflation.  As such, we believe further investigation
of it is necessary.

In regards the potential implications of 
the results discussed in this talk to 
inflationary cosmology, 
we infer that under generic circumstances the scalar inflaton field 
will dissipate a nonnegligible amount of radiation during 
inflation.  In particular,  
the lower bound suggested by the above estimates  
already are sufficiently 
high to preclude a mandatory requirement 
for a reheating period.
Moreover, the high temperature results of case (b) could lead
to robust warm inflation.  These results suggest that
warm inflation may be a very natural dynamics once proper
treatment of the dissipative dynamics is made,
thus corroborating with similar suggestions from
other considerations.
For one, as mentioned above, the quantum-classical
problem is greatly alleviated in warm inflation.
Second, as shown in \cite{bg}, a dissipative component in the scalar field
field evolution equation can alleviate the initial
condition problem of inflation, since dissipative effects
will damp kinetic energy, thus allow the potential
energy to dominate more quickly.  Third,
as an interesting addition, presence
of radiation during inflation makes it conducive to
generate large scale magnetic fields of sizable
magnitudes \cite{bkw}.
Finally and perhaps most important, as already noted
the presence of radiation generally affects the seeds of
density perturbations.  Up to now only simplified models
based on ideal thermalization have been considered
\cite{bf2,abden,wimodels}.  Beyond that, much remains
to be understood about the role radiation plays in
influencing density fluctuations.  One of the hopes
of warm inflation is that it may solve the scalar
potential fine tuning problem, but a qualified attempt at
this requires much further understanding about the
interplay between radiation and the scalar field.
As a hypothetical example to illustrate the importance of
this problem,  if one went to the opposite limit
from ideal thermalization to that of negligible influence from
the radiation field on the density perturbations, so
that only the quantum fluctuations contributed, then
interestingly warm inflation models would {\it not}
require fine tuning, so that scalar potentials of
reasonably large curvature would be acceptable.
This is a revealing observation, which provides motivation
to better understand the full dynamics of field interactions
in warm inflation models.

In summary, warm inflation shows promise as a dynamical solution 
to the cosmological puzzles.  In regards the crucial question
of its first principles quantum field theory dynamics,
verification of the
expectations discussed above requires a proper extension of these
calculations to expanding spacetime, and within a nonequilibrium
formulation \cite{lawrie2}, which we plan to examine.

\end{document}